\documentclass[%
reprint,
 amsmath,amssymb,
 aps,showkeys,
prl
]{revtex4-1}

\usepackage{graphicx}
\usepackage{dcolumn}
\usepackage{bm}
\usepackage{hyperref}
\usepackage{placeins}
\usepackage{color}
\usepackage{subcaption}

\begin{document}
\title{Optimal Measurement Protocols in Quantum Zeno Effect}

\author{Sergey Belan}
\email{sergb27@yandex.ru}
\affiliation{Massachusetts Institute of Technology, Department of Physics, Cambridge,
Massachusetts 02139, USA}
\author{Vladimir Parfenyev}
\affiliation{Landau Institute for Theoretical Physics,
 142432 Chernogolovka, Russia}

 \begin{abstract}
The quantum Zeno effect is the prediction, going back to Alan Turing, that the decay of an unstable system can be slowed down by measuring it frequently enough. It was also noticed later that the opposite effect, i.e., enhancement of the decay due to frequent measurements, is rather common.
An important question arising in this regards is how to choose the optimal measurement strategy to achieve the highest possible decay rate.
Here we rigorously prove the universal optimality of the stroboscopic measurement protocol in the sense that it always provides the shortest expected decay time among all measurement procedures. However, the implementation of the stroboscopic protocol requires the knowledge of the optimal sampling period which may depend on the fine details of the quantum problem. We show that this difficulty can be overcame with the simple non-regular measurement procedures inspired by the scale-free restart strategies used to speed up the completion of random search tasks and probabilistic algorithms in computer science.
Besides, our analysis reveals the universal criteria to discriminate between Zeno and anti-Zeno regimes for the unstable quantum systems under Poissonian measurements.
 \end{abstract}

\maketitle

As was recognized soon after establishing the foundations of the quantum theory, the measurements can slow down the evolution of the unstable quantum systems.
In the idealized limit of infinitely frequent instantaneous measurements the decay is completely suppressed so that the system is frozen to the initial excited state.
This phenomenon had first been qualitatively described by Alan Turing \cite{Hodges} and is named Zeno effect \cite{Misra_1977} - after Greek philosopher Zeno of Elea famous for his paradoxes challenging the logical possibility of motion.

During past decades the measurement-induced slow down of quantum evolution has been investigated for a variety of experimental setups \cite{Itano_1990,Kwiat_1995,Kakuyanagi_2015,Slichter_2016,Streed_2006,Helmer_2009,Wolters_2013,Patil_2015,Kalb_2016}.
Importantly, the opposite phenomenon - the quantum anti-Zeno effect or inverse Zeno effect \cite{Kofman_2000, Facchi_2001} - has also been demonstrated experimentally \cite{Fischer_2001, Harrington_2017}.
This open the door to the opportunity to optimize the quantum decay by the carefully tuned schedule of measurements.
However, given a myriad of possible combinations of underlying quantum evolution and measurement protocols, it remains unclear if  there are any general guiding principles to design the optimal measurement procedure providing the highest possible decay acceleration for a given quantum system.


To address this challenge, here we develop a general theoretical approach to quantum Zeno effect for a generic quantum system under an arbitrary measurement protocol assuming instantaneous and ideal measurements.
Our analysis focuses on the behaviour of the average time required to detect the decay in the series of successive measurements.
This metric exhibits a non-monotonic behaviour in dependence on the measurement frequency so that one can achieve the optimal decay conditions by bringing the system to the point of Zeno/anti-Zeno transition.
Surprisingly, we found that the optimal detection strategy for any system is very simple: one should apply measurements in a strictly periodic manner, i.e. each $\tau_\ast$ units of time, where $\tau_\ast$ depends on the details of the system under study.
Since the optimal measurement period of such a stroboscopic protocol is determined by the parameters of the quantum system, which may be poorly specified or unknown a priory, we also propose the non-uniform measurement protocols whose performance is weakly sensitive to such details.
These protocols resemble the restart strategies previously proposed to improve the mean completion time of the probabilistic algorithms in computer science \cite{Luby_1993} and random search tasks \cite{Kusmierz_2018,Belan_2019}.
Finally, we discuss the practically important example of randomly distributed Poissonian measurement events, which is sometimes dictated by the experimental technique \cite{Power_1996,Belavkin_2000,Shushin_2011}, and reveal universality exhibited by the quantum systems in the point of Zeno/anti-Zeno transition.
This also allows us to formulate the simple conditions to discriminate between Zeno and anti-Zeno regimes.
All these findings are illustrated with the example of Zeno effect in a system of cold atoms \cite{Wilkinson_1997, Fischer_2001}.

\textbf{Results}

\textbf{A theoretical framework.}
In the quantum description of an unstable system a key role is played by the survival probability $P(T)$, i.e. the probability of finding the system in its initial unstable state after time $T$ in the absence of any measurements performed on the system during this time. The dynamic of the unstable system is irreversible: the initial unstable state decays with a finite lifetime, and the system never returns to the initial state spontaneously. Such an irreversible dynamics takes place when the initial state is coupled to the environment whose continuous states have energies in a wide range.

We are interested in the scenario, when the unstable system undergoes a series of measurements during the time evolution to check whether it is still in its initial unstable state. The measurement protocol is characterised by the sequence of the inter-measurement time intervals $\{T_k\}_{k=1}^{+\infty}=T_1,T_2,\dots$. Importantly, we treat the measurements as instantaneous projections and assume that the survival probability after several measurements factorizes. This is justified when the time required to perform a measurement is small compared to inter-measurement interval and when the system-environment correlations can be neglected.

The previous theoretical studies of Zeno effect focused mainly on the calculation of the effective decay rate extracted from the behaviour of the measurements-modified survival probability, see, e.g., \cite{Koshino_2005,Chaudhry_2014,Chaudhry_2016,Aftab_2017,Chaudhry_2017,Zhang_2018}.
In contrast, here we consider the mean detection time as the main metric of interest.
Namely, we calculate the expectation of the random time $t$ when the measurement attempts finally result in detection of decay.
It is straightforward to show that the mean decay time defined in this way is equal to the inverse effective decay rate provided the deviation of $P(T)$ from unity for the typical inter-measurement interval is small (see Methods).
Thus, these two approaches to quantification of Zeno effect becomes equivalent in the limit of sufficiently frequent measurements.
Note also, that the mean detection time is more natural and well-defined metric while dealing with the non-uniform measurement protocols discussed below.

The random time required to detect that system has decayed can be determined from the following infinite chain of equations
\begin{eqnarray*}
t=T_1+t_1x_{T_1},\\
t_1=T_2+t_2x_{T_2},\\
...\\\
t_n=T_{n+1}+t_{n+1}x_{T_{n+1}},\\
...
\end{eqnarray*}
where $x_{T_k}$ is the binary random variable which is equal to zero if the outcome of the $k$-th measurement is "decay" and is unity otherwise, and $t_i$ represents the time remaining to the decay provided the $i$-th measurement was "not decayed".
Intuition behind this set of equations is very simple: if the next measurement attempt failed to detect decay, the system quantum evolution starts anew from the initial (undecayed) state.
Performing averaging over the quantum statistics and taking into account that $\langle x_{T_k}\rangle=P(T_k)$  one obtains
\begin{equation}
\label{decay_time}
\langle t\rangle = T_1 + \sum_{n=2}^{\infty} T_n\prod_{k=1}^{n-1}P(T_k).
\end{equation}
The later equation relates the expectation of the detection time to the quantum survival probability of the measurement-free system. Once $P(t)$ is known, Eq. (\ref{decay_time}) allows to calculate the expected decay time for any sequence of the inter-measurement intervals $\{T_k\}_{k=1}^{+\infty}$.

\begin{figure}[t]
  \includegraphics[width=0.85\linewidth]{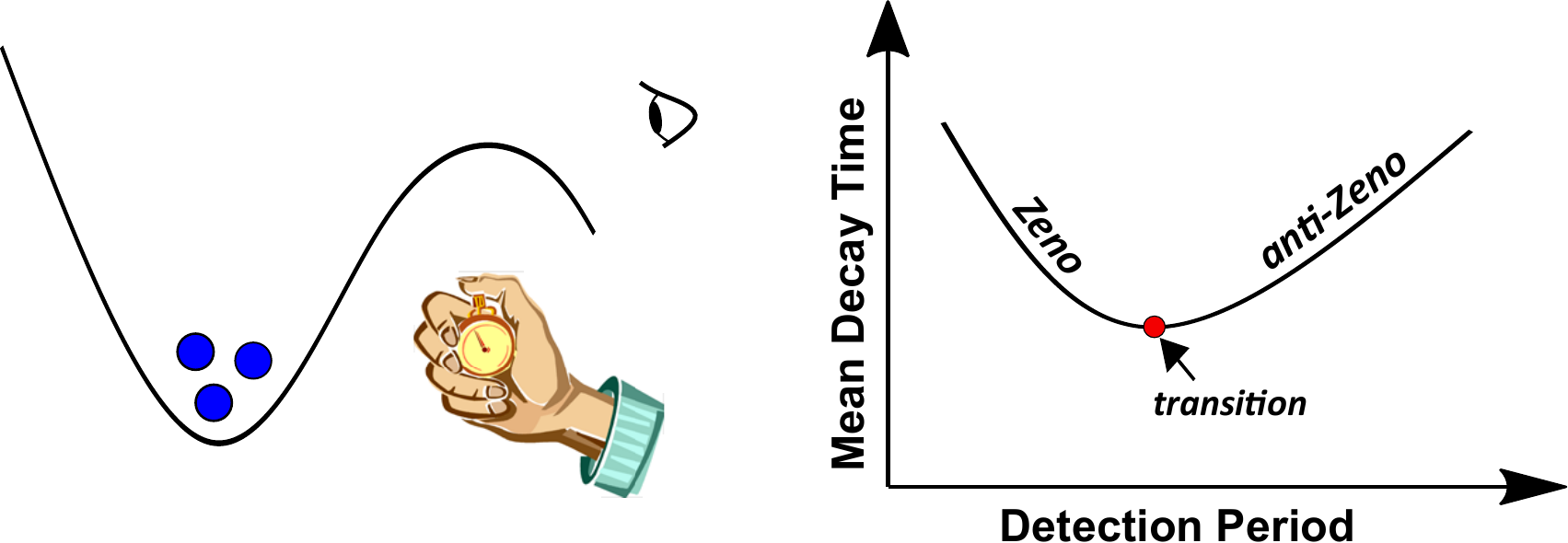}
  \captionsetup{format=plain, justification=centerlast}
  \caption{The sketch of the measurement of the time to decay in quantum tunnelling experiment (left). The expected time required to detect the decay in a series of stroboscopic measurements attains a minimum corresponding to the Zeno-anti-Zeno transition (right).}
\label{fig:0}
\end{figure}

Let us discuss how this formula simplifies for the particular measurement strategies most studied in the existing literature.
In the original formulation of quantum Zeno effect \cite{Misra_1977}, the measurement protocol was stroboscopic, i.e. the  measurement events are equally spaced in time, $\{T_k\}_{k=1}^{+\infty}=\tau,\tau,\dots$.
In this case, the right-hand-side of Eq. (\ref{decay_time}) represents the sum of the infinite geometric series with the common ratio $P(\tau)$ so that we readily find
\begin{equation}
\label{stroboscopic}
\langle t_{\tau}\rangle=\frac{\tau}{1-P(\tau)}.
\end{equation}
Note that at the very beginning of the decay $1 - P(\tau) \propto \tau^2$ \cite{Ghirardi_1979} and therefore $\langle t_{\tau}\rangle \propto 1/\tau$, whereas for the very large detection period $\langle t_{\tau}\rangle \propto \tau$. Thus, the dependence of mean decay time on the stroboscopic period always attains minimum, see Fig.~\ref{fig:0}.

Another important scenario is the stochastic measurements protocol where measurement events are separated by random time intervals \cite{Power_1996,Belavkin_2000,Shushin_2011,Gherardini_2016,Muller_2017}.
In this case, $T_i$'s are assumed to be independent and identically distributed random variables sampled from the probability density $\rho(T)$ with the well-defined first moment $\langle T\rangle=\int_0^\infty dT \rho(T)T$.
To calculate the expected decay time, we should additionally average Eq. (\ref{decay_time}) over the statistics of inter-measurement intervals.
This gives
\begin{equation}
\label{stochastic}
\langle t\rangle=\frac{\langle T \rangle}{1-\langle P(T) \rangle}=\frac{\int_0^\infty dT \rho(T)T}{1-\int_0^\infty dT \rho(T)P(T)}.
\end{equation}
Clearly, the stroboscopic protocol with period $\tau$ can be treated as a particular case of randomly distributed protocol having  measurement intervals distribution $\rho(T)=\delta(T-\tau)$.


The behaviour of $\langle t\rangle$ as a function of $\langle T\rangle$ allows us to identify the Zeno and anti-Zeno regimes.
Namely, expected decay time can either increase
with decreasing $\langle T\rangle$, which we refer to as the Zeno effect or
decrease for more frequent measurements, which we define
as the anti-Zeno effect.
Equivalent definition of Zeno and anti-Zeno effects in terms of effective decay rate was adopted, e.g., in Refs. \cite{Kofman_2000,Segal_2007,Thilagam_2010,Chaudhry_2014,Chaudhry_2016}.
As we will see below, in general $\langle t\rangle$ in its dependence on $\langle T\rangle$ exhibits minimum corresponding to Zeno/anti-Zeno transition.



\textbf{Optimality of stroboscopic measurement protocol.}
For the sake of illustration, let us consider a system consisting of ultra-cold atoms that are trapped in an accelerating periodic optical potential of the form $V_0 \cos(2 k_L x - k_L a t^2)$, where $V_0$ is the potential amplitude, $k_L$ is the laser wavenumber, $x$ is position in the laboratory frame, $a$ is the acceleration and $t$ is time -- the experimental setup intensively discussed in Refs.~\cite{Wilkinson_1997, Fischer_2001}. In the accelerating reference frame, $x' = x - at^2/2$, this potential becomes $V_0 \cos(2k_L x') + M a x'$, where $M$ is the mass of atoms and the last term corresponds to the inertial force experienced by them. The energy spectrum of the considered system under the assumption of zero acceleration consists of Bloch bands separated by gaps. As an initial condition, we assume that the lowest band is uniformly occupied, while the higher bands are empty. When an acceleration is imposed, the atomic quasimomentum changes as the atoms undergo Bloch oscillations. The atoms can escape from the accelerating lattice by interband Landau-Zener tunneling and hence form an unstable quantum system. Tunneling between Bloch bands should not be confused with the totally different process of tunneling between different wells of the periodic potential.

The escape probability for trapped atoms can be analytically calculated and it exhibits deviation from exponential decay for short times, see Ref.~\cite{Wilkinson_1997} and Methods. The experimental technique intended to repeatedly measure the amount of trapped atoms was developed in Ref.~\cite{Fischer_2001} and it was shown that the expression for the survival probability (\ref{eq:surv_prob}) together with the assumption of instant and ideal measurements lead to the results that are in a good agreement with the experiment. For compactness, below we will measure the time $t$ in units of $M/(4 \hbar k_L^2)$, the potential amplitude $V_0$ in $4 \hbar^2 k_L^2/M$, and acceleration $a$ in $8 \hbar^2 k_L^3/M^2$.

\begin{figure}
\begin{subfigure}[h]{0.49\linewidth}
\includegraphics[width=\linewidth]{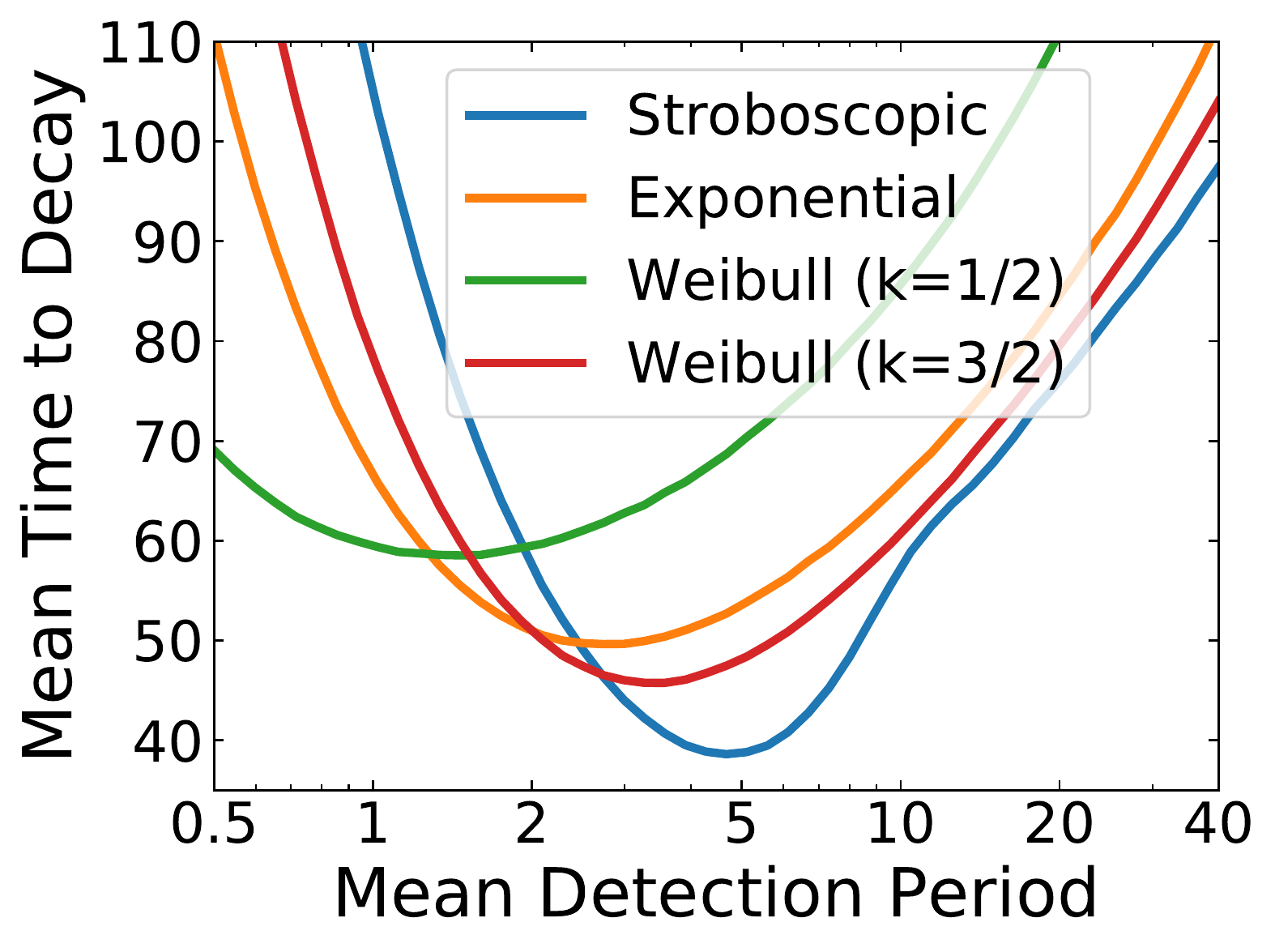}
\end{subfigure}
\begin{subfigure}[h]{0.49\linewidth}
\includegraphics[width=\linewidth]{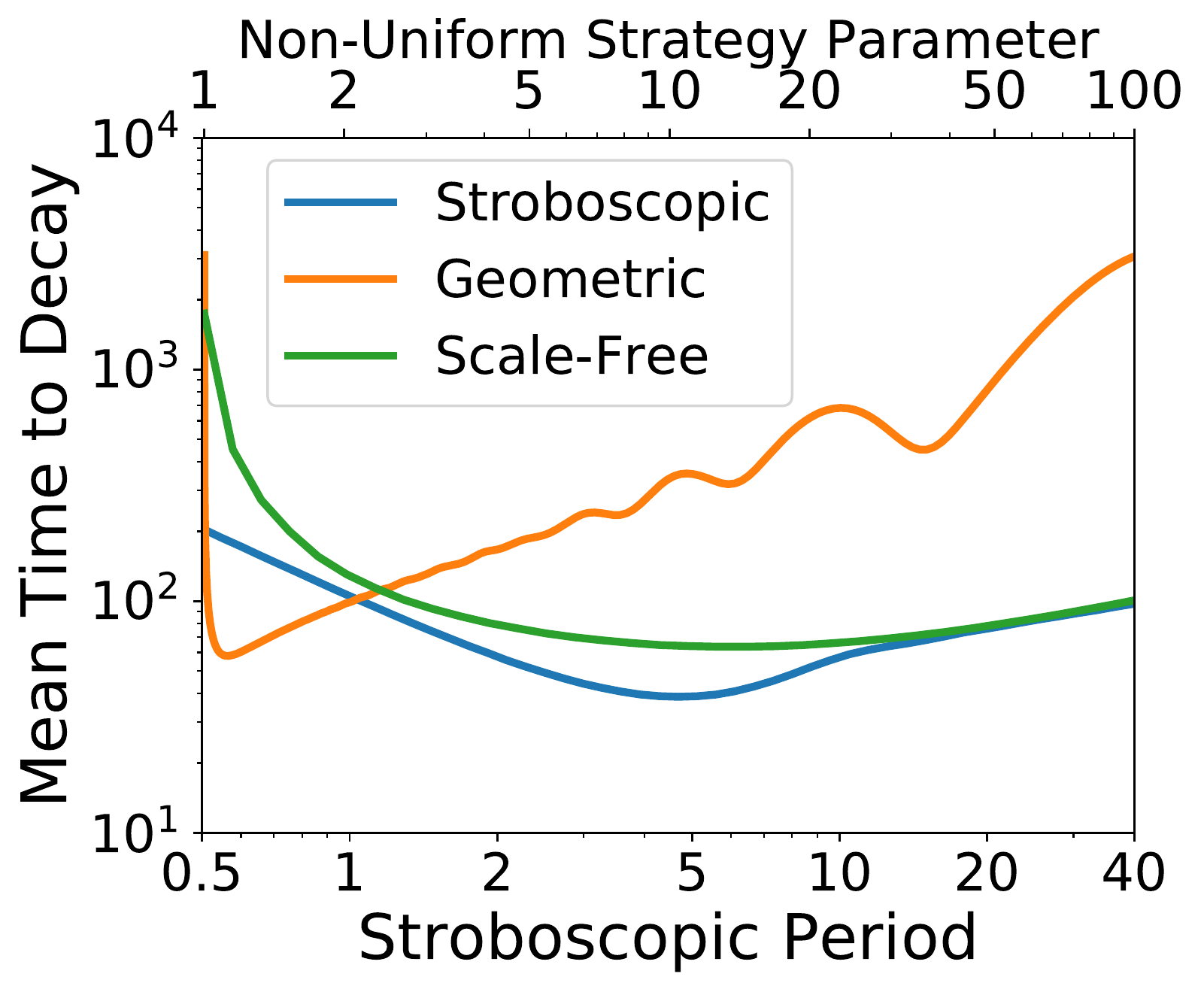}
\end{subfigure}%
\captionsetup{format=plain, justification=centerlast}
\caption{(a) The mean time to decay $\langle t \rangle$ as a function of mean detection period $\langle T\rangle$ for various uniform in time stochastic measurement strategies. (b) The stroboscopic protocol also outperforms non-uniform in time geometric and scale-free measurement protocols (see section Results for the definition of these measurement strategies). The survival probability is taken for $a=0.1$ and $V_0 = 0.4$, the initial time-step for geometric protocol is $\tau_0 = 0.01$.}
\label{fig:23}
\end{figure}

In Fig.~\ref{fig:23}a we plot the expected tunnelling time as a function of mean inter-measurement intervals $\langle T\rangle$ for various stochastic measurement strategies having uniform in time statistics.
We see that a minimum of $\langle t\rangle$ is always achieved and
while the values taken by the different minima and their positions depend on the distribution of the inter-measurement time, it is stroboscopic protocol that provides the lowest of minima.
Also, Fig.~\ref{fig:23}b demonstrates that stroboscopic protocol outperforms the non-uniform in time measurement procedures.

These observations allow us to conjecture that in the case of ultra-cold atoms, stroboscopic protocol is the optimal measurement strategy. Surprisingly, this is also true in general. In Methods we rigorously prove that for any quantum system the deterministic measurement strategy is optimal among all possible strategies. Namely, our result is the following: if we find an optimal period $\tau_\ast$ of stroboscopic measurements such that the expected decay time a quantum system under study as a function of $\tau$ attains a global minimum $\langle t_{\tau_\ast}\rangle$, then this performance cannot be beaten by any other measurement procedure.

\textbf{Luby-like measurement protocol.}
As we learn from the above consideration, to achieve the maximal performance one should implement the stroboscopic measurements with carefully tuned sampling period $\tau_\ast$ which depends on the details of the underlying quantum process.
An interesting question naturally emerges: is it possible to have realizations of
the measurement sequence that give values of the expected decay time close to the optimal without detailed knowledge of the system parameters?
Previously, a similar challenge motivated the development of the restart strategies improving the mean completion time of probabilistic algorithms and Internet tasks without introducing any knowledge of the underlying statistics~\cite{Luby_1993,Kusmierz_2018, Belan_2019}.
Inspired by the appealing analogy between restart and projective measurement (indeed, the projection postulate tells us that at every moment one confirms the survival of the system through the measurement, the quantum state of the system is reset to the initial undecayed one and its evolution starts from scratch), we investigate the effect of the Luby-like protocol \cite{Luby_1993} on the expected decay time.
Namely, we assume that the inter-measurements intervals are given by $\{T_k\}_{k=1}^{+\infty}=\{\tau_0$, $\tau_0$, $2\tau_0$, $\tau_0$, $\tau_0$, $2\tau_0$, $4\tau_0$, $\tau_0$, $\tau_0$, $2\tau_0$, $\tau_0$, $\tau_0$, $2\tau_0$, $4\tau_0$, $8\tau_0$, $\tau_0$, $\dots \}$. The $i$-th term in this sequence is
\begin{equation}
\label{Luby}
T_i=\left\{\begin{array}{ll}
2^{k-1}\tau_0 ,\ \ \text{if}\ \  i=2^{k}-1,\\
\\
T_{i-2^{k-1}+1},\ \ \text{if} \ \  2^{k-1}\le i<2^{k}-1,
\end{array} \right.
\end{equation}
where the initial time step $\tau_0$ should be smaller then the expected decay time of the system under study.


This strategy possess the following remarkable property: the total
time spent on inter-measurement runs of each length is roughly equal.
This is because each time a pair
of runs of a given length has been completed, a run of
twice that length is immediately executed.
Due to this property, the strategy does not introduce any characteristic time scale into the measurement process.

\begin{figure}
\begin{subfigure}[h]{0.49\linewidth}
\includegraphics[width=\linewidth]{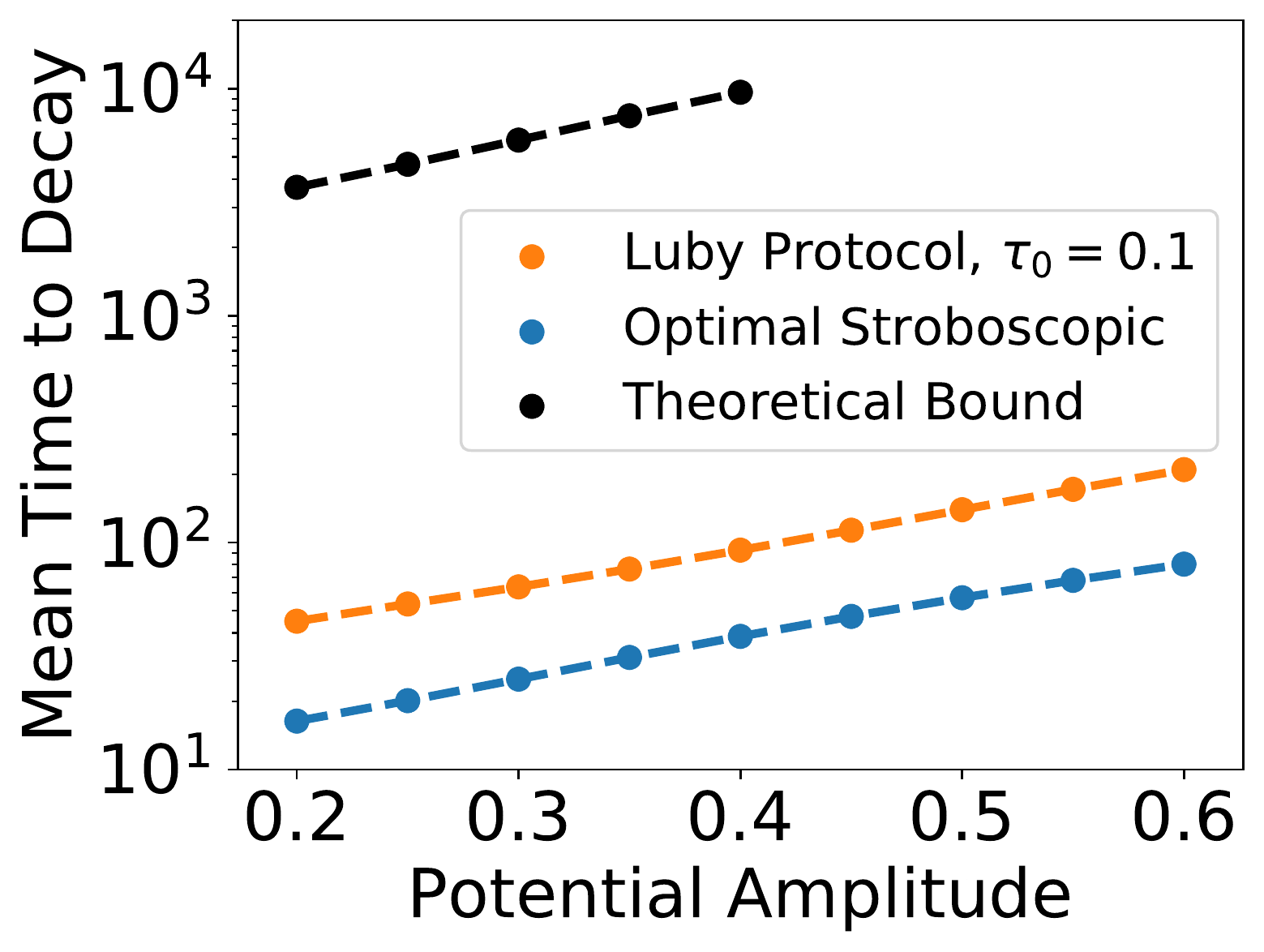}
\end{subfigure}
\begin{subfigure}[h]{0.49\linewidth}
\includegraphics[width=\linewidth]{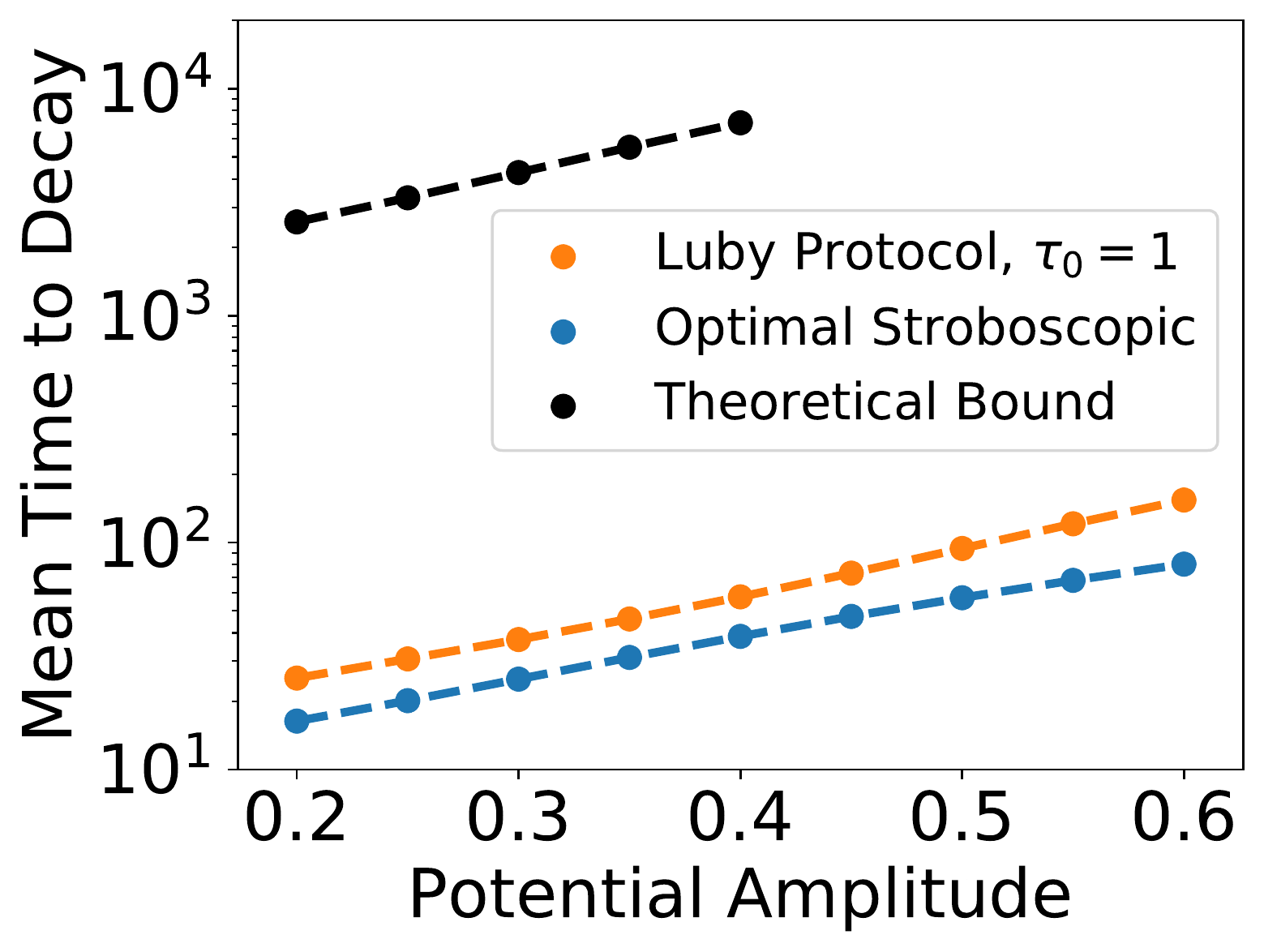}
\end{subfigure}%
\captionsetup{format=plain, justification=centerlast}
\caption{The comparison of Luby measurement strategy with the optimal stroboscopic protocol for different values of the potential amplitude. The acceleration of optical potential is equal to $a=0.1$. The black dashed line is defined only for the sufficiently small values of potential amplitude, since larger $V_0$ lead to non-monotonic behaviour of the survival probability $P(T)$, thus, violating the assumptions underlying the proof of Eq. (\ref{Luby_bound}).}
\label{fig:45}
\end{figure}

It turns out that Luby strategy allows to come surprisingly close to the optimum value provided by the best stroboscopic protocol.
Namely, in Methods we derive the following inequality
\begin{equation}
\label{Luby_bound}
\langle t_{univ} \rangle \le C_1 \langle t_{\tau_\ast}\rangle (\log_2 \frac{\langle t_{\tau_\ast}\rangle}{\tau_0}+C_2),
\end{equation}
which is universally valid for any quantum system whose survival probability $P(t)$ is a monotonically decreasing function. Here $C_1=\frac{8}{(1-e^{-1})^2}$ and $C_2=(1-e^{-1})^2\sum_{k=1}^\infty ke^{-k+1}\log_2 k+3$. Thus, the decay of an unstable quantum system under Luby measurement strategy is only a logarithmic factor slower than the decay of this system subject to the optimally tuned stroboscopic measurements.
 Note that in practice the performance of the Luby protocol is much better that the upper bound predicted by Eq. (\ref{Luby_bound}), see Fig.~\ref{fig:45}.
 This is because in derivation of Eq. (\ref{Luby_bound}) we were struggling mainly for the logarithmic factor and did not tried to reduce the numerical constants $C_1$ and $C_2$ in the involved estimates.

\textbf{Other scale-free strategies.}
Next we introduce a class of non-uniform measurement protocols inspired by the scale-free restart strategies developed to speed-up the completion of random search tasks \cite{Kusmierz_2018,Belan_2019}.
These strategies are effective when the survival probability of the quantum system under study has the form $P(T)=f(T/T_0)$, where $T_0$ is the characteristic time associated with quantum evolution.

\begin{figure}
\begin{subfigure}[h]{0.49\linewidth}
\includegraphics[width=\linewidth]{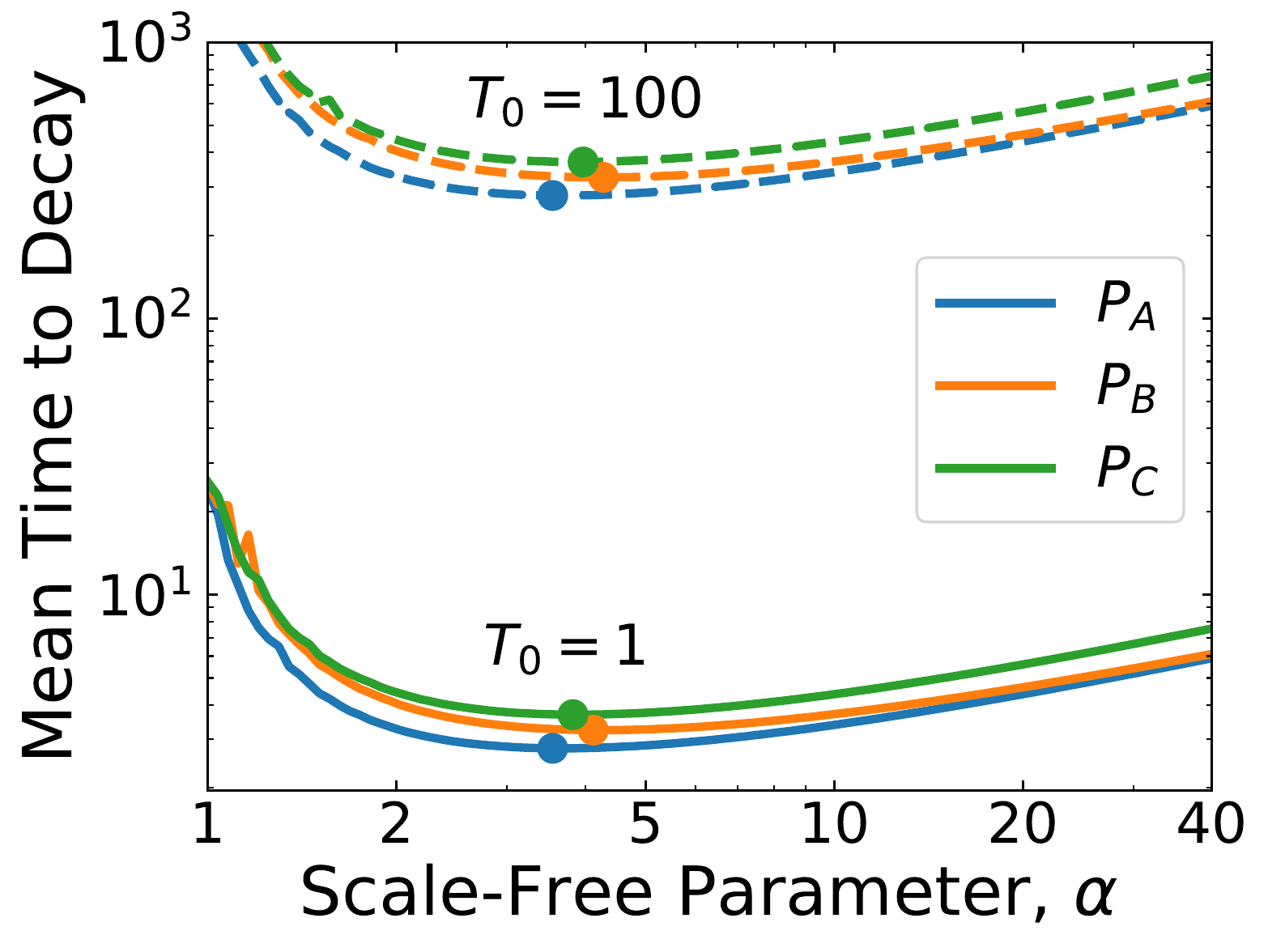}
\end{subfigure}
\begin{subfigure}[h]{0.49\linewidth}
\includegraphics[width=\linewidth]{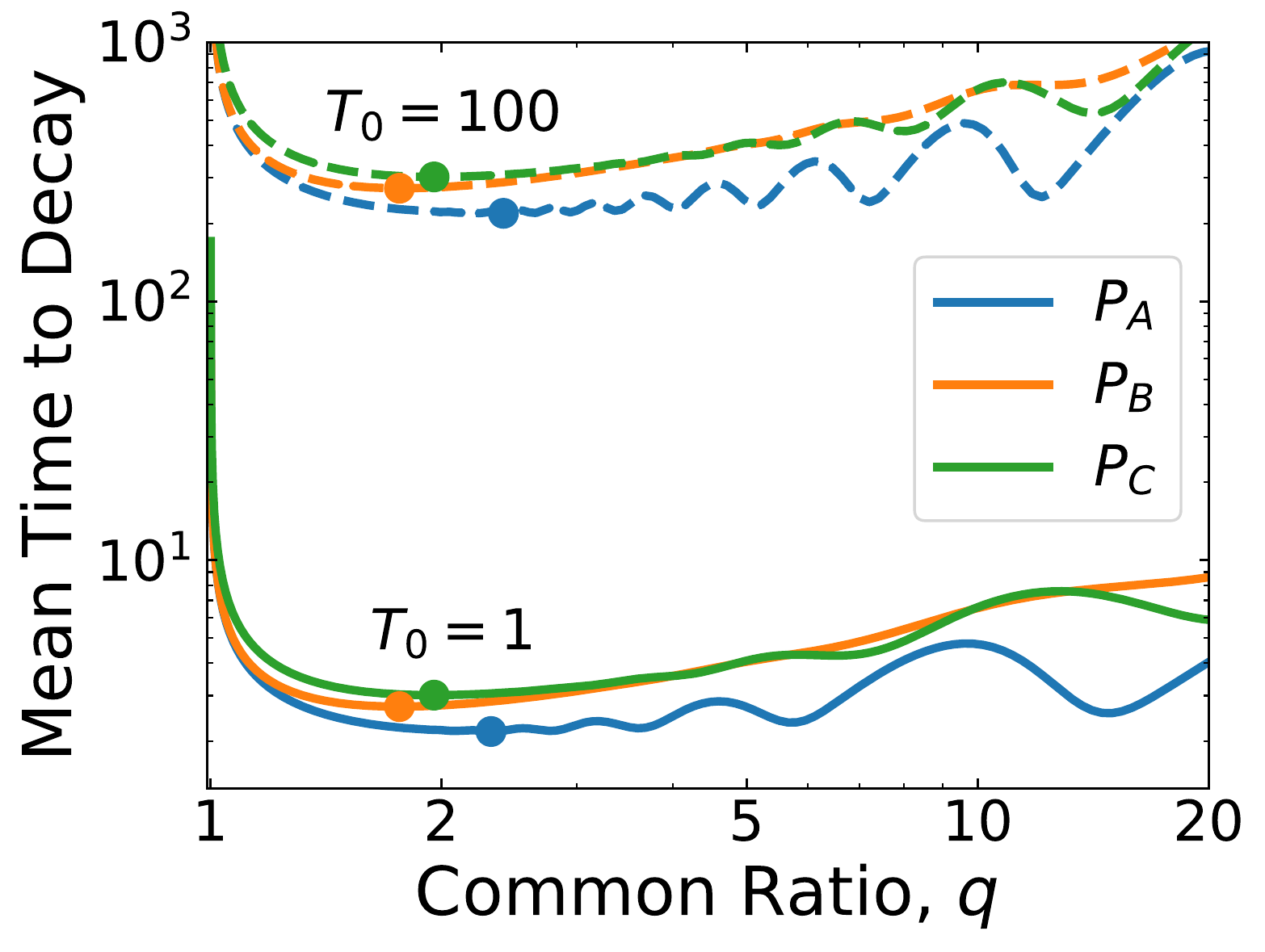}
\end{subfigure}%
\captionsetup{format=plain, justification=centerlast}
\caption{Demonstration that the optimal values of strategy parameters do not depend on the characteristic time scale $T_0$ for scale-free (a) and geometric (b) protocols, $\tau_0 = 0.01$. The involved survival probabilities are $P_A = \exp\left[-(T/T_0)^2\right]$, $P_B=1/\left[1+(T/T_0)^2\right]$, and $P_C= \left[ 1 + \frac{T/T_0}{1+(T/T_0)^2}\right] \exp[-T/T_0]$.}
\label{fig:67}
\end{figure}

Assume that the measurements are applied at random time moments at rate which is inversely proportional to the time elapsed since the start of the experiment, i.e. $r(t)=\alpha/t$, where $\alpha$ is a dimensionless constant.
As we see in Fig. \ref{fig:67}a, such measurement protocol does not need to be optimized with respect to the characteristic time scale $T_0$ of the underlying quantum problem:  the mean decay time as a function of $\alpha$ attains its minimum at optimal value $\alpha_\ast$ which is determined by the particular form of the function $f(T/T_0)$, but does not depend on the time scale $T_0$ (see Methods for the rigorous proof).
The same property is exhibited by the geometric protocol~\cite{Belan_2019}, see Fig. \ref{fig:67}b.
Here the measurement time instants are chosen from the geometric sequence $\{T_k\}_{k=1}^{+\infty}=\tau_0,q\tau_0,q^2\tau_0,...$, where $\tau_0\ll T_0$ is the initial time-step and $q$ is the dimensionless common ratio.
Again the optimal value $q_\ast$ minimizing the expected decay time does not depend on the scale $T_0$.
Thus, once the form of the survival probability $f(T/T_0)$ is known, one can optimize the system decay using the scale-free sampling of the measurement times even in the absence of reliable estimate of the characteristic time scale $T_0$.

An important open question, which is beyond of the scope of the current article, is if it is possible to derive
the rigorous bounds on the performance of the stochastic scale-free and geometric
measurement protocols similarly as it was done above for
the Luby protocol.

\textbf{Universality in statistics of decay under optimal Poissonian measurements.}
Sometimes, the form of the measurement protocol is dictated by the experimental conditions.
Say, in some experiments, the measurement process is of stochastic nature so that the time interval between two consecutive measurements is randomly varied \cite{Power_1996,Belavkin_2000,Shushin_2011}.
A particularly important example is the Poisson measurements for which the inter-measurement intervals are sampled from the exponential distribution, i.e. $\rho(T)=re^{-rT}$, where $r$ is the measurements rate.
In this subsection we reveal the universal feature associated with the optimally adjusted Poisson measurements.

In this case Eq. (\ref{stochastic}) reduces to
\begin{equation}
\label{Poisson}
\langle t_r\rangle=\frac{1}{r(1-r\tilde{P}(r))}
\end{equation}
where $\tilde{P}(r)$ is the Laplace transform of $P(t)$ evaluated at $r$.
It is also straightforward to arrive at the following result for the second moment of the decay time (see Methods)
\begin{equation}
\label{Poisson2}
\langle t_r^2\rangle=\frac{2(1-r\tilde{P}(r)-r^2\partial_r \tilde{P}(r))}{r^2(1-r\tilde{P}(r))^2}.
\end{equation}
Now let us assume that we found the optimal measurement rate $r_\ast$ such that the decay rate attains its (local or global) maximum.
Then $\partial_r \langle t_{r_\ast}\rangle=0$ and Eq.~(\ref{Poisson2}) becomes
\begin{equation}
\label{Poisson_universality}
\langle t_{r_\ast}^2\rangle=2\langle t_{r_\ast}\rangle^2-2r_\ast^{-1}\langle t_{r_\ast}\rangle.
\end{equation}
Equation (\ref{Poisson_universality}) is a universal property common to all quantum systems subject to optimally tuned Poissonian measurements.
A remarkable consequence of Eq. (\ref{Poisson_universality}) is the inequality relating the optimal measurement rate and the resulting expectation of the decay time.
Indeed, since $\langle t_{r_\ast}^2\rangle\ge \langle t_{r_\ast}\rangle^2$, we immediately find that  $\langle t_{r_\ast}\rangle \ge 2r_\ast^{-1}$.

The point of Zeno-anti-Zeno transition for Poissonian measurements is also characterized by another universal equality (see Methods)
\begin{equation}
\label{Poisson_universality_2}
r_\ast\tilde{T}_{r_\ast}=2,
\end{equation}
where the time $\tilde{T}_{r}=\frac{\int_0^\infty dT e^{-rT}T(1-P(T))}{\int_0^\infty dT e^{-rT}(1-P(T))}$ has the following meaning.
Assume that we apply a single projective measurement at random (Poisson) moment of time.
Then $\tilde{T}_{r}$ is the average time of those trials that gave the outcome "decayed".

\begin{figure}[t]
  \includegraphics[width=0.65\linewidth]{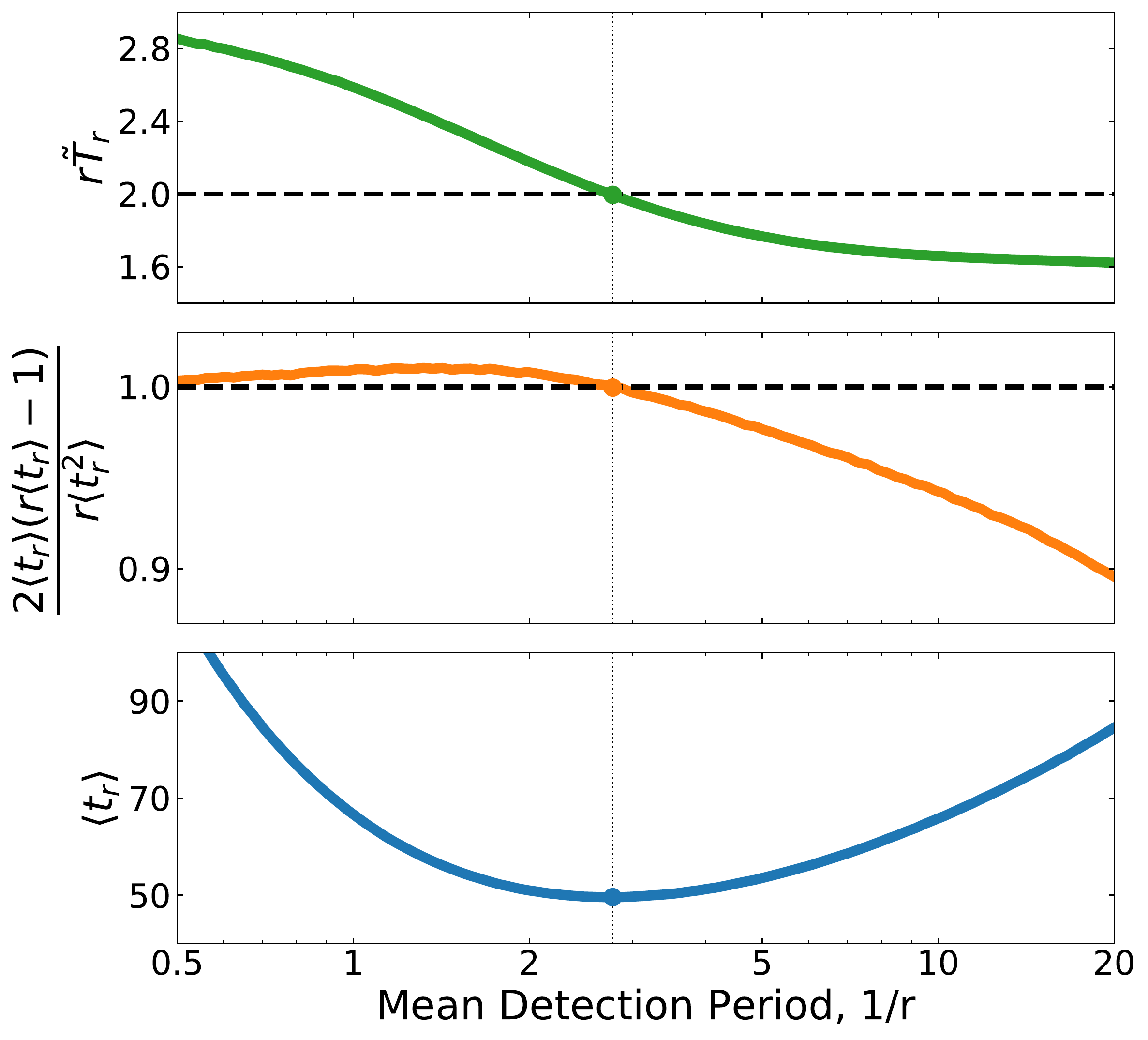}
  \captionsetup{format=plain, justification=centerlast}
  \caption{Illustration of the universal features associated with the optimally adjusted Poisson measurements.}
\label{fig:8}
\end{figure}

The universal identities revealed above allows us to formulate the simple and practically important criteria to distinguish between the quantum Zeno and the anti-Zeno effects.
Namely, it is straightforward to show that the conditions $\langle t_{r_\ast}^2\rangle<2\langle t_{r_\ast}\rangle^2-2r_\ast^{-1}\langle t_{r_\ast}\rangle$ (or $r\tilde{T}_{r}>2$) and $\langle t_{r_\ast}^2\rangle> 2\langle t_{r_\ast}\rangle^2-2r_\ast^{-1}\langle t_{r_\ast}\rangle$ (or $r\tilde{T}_{r}<2$) indicate the Zeno and anti-Zeno regimes, respectively (see Fig. \ref{fig:8} for the illustration).
Let us stress that deviations from Eqs. (\ref{Poisson_universality}) and (\ref{Poisson_universality_2}) may occur when the inter-measurement intervals are not taken from the
exponential distribution and, for this reason, these criteria are universally valid only in the case of Poisson measurements.

\textbf{Discussion}

The quantum Zeno effect is an intriguing topic attracting ongoing interest from both experimental \cite{Itano_1990,Kwiat_1995,Kakuyanagi_2015,Slichter_2016,Streed_2006,Helmer_2009,Wolters_2013,Patil_2015,Kalb_2016,Fischer_2001, Harrington_2017} and theoretical \cite{Kofman_2000,Facchi_2001,Koshino_2005,Chaudhry_2014,Chaudhry_2016,Aftab_2017,Chaudhry_2017,Zhang_2018,Barone_2004,Chen_2010} sides.
In contrast to majority of previous studies aimed to explore the peculiarities of Zeno dynamics in particular classes of systems, here we adopt the general model-independent approach which allowed us to address theoretically the optimization questions and the issue of universality.

Our analysis proves that the regular stroboscopic sampling -- the most popular measurement protocol in the quantum Zeno research literature -- is a universally optimal strategy providing
the best performance (i.e. the lowest expected decay time in the point of Zeno/anti-Zeno transition) for any given quantum system.
However, in general, the finding of the optimal sampling rate is not a simple task.
Here we proposed the computer science inspired solution to this problem.
Namely, by noting the formal similarity between the projective measurement of a quantum system and the restart of a classical stochastic process \cite{Luby_1993,Montanari_2002, EM_2011,Evans_2011,Evans_2014,Kusmierz_2014,Kusmierz_2015,Rotbart_2015,Pal_2016,Eule_2016,Reuveni_PRL_2016,Nagar_2016,
Roldan_PRE_2017,Reuveni_PRL_2017, Belan_2018, Kusmierz_2018, Chechkin_2018, Pal_2019, Belan_2019}, we demonstrated that detection protocols reminiscent to the scale-free restart strategies used to speed up progress of randomized tasks help to achieve near-to-optimal decay rate without detailed knowledge of the system parameters.

Previously, considerable theoretical efforts have been made to find the conditions to discriminate between the quantum Zeno and the anti-Zeno effects.
Say, according to Refs. \cite{Facchi_2001, Koshino_2005,Facchi_2008}
one should scrutinize the general features of the system-environment
interaction and calculate the residue of the propagator involving the environmental spectral density function.
In Ref.~\cite{Zhang_2018} an explicit analytical criterion is derived which relates the sign of Zeno dynamics of the two-level system in a dissipative environment with the convexity of the spectrum.
In contrast, our approach gives the criteria in terms of the directly measurable statistics of the decay time and does not refer to any details of the interaction Hamiltonian, thus, providing an appealing practical tool to probe Zeno and anti-Zeno regimes.

Besides the issue of optimal control of the unstable quantum systems, another promising application of the above theory lies in the fields of quantum walks and quantum search algorithms.
Assume that the quantum particle undergoes a unitary evolution and the series of measurements are performed (either in regular or in random fashion) to see if the particle has reached the target region of its phase space.
What is then the expected time required to detect the first arrival of the particle?
This question is known as quantum first detection problem \cite{Krovi_2006,Varbanov_2008,Stefanak_2008,Darazs_2010,Dhar_2015a,Dhar_2015b,Friedman_2017,Thiel_2018a,Thiel_2018b,Lahiri_2019}, which is of basic notion for design of quantum search algorithms.
As first noticed in Ref. \cite{Varbanov_2008}, in some cases the optimum sampling period exists that bring the mean first detection time to the minimum.
We anticipate that the approach presented here may help to uncover the universal aspects of this kind of optimal behaviour.


\textbf{Methods}

\textbf{Effective decay rate.} For the stroboscopic measurement protocol with period $\tau$, the probability of finding the system in its initial state after $N$ measurements (i.e. after time $t=N\tau$) is equal to $P(\tau)^N=e^{-\Gamma t}$, where $\Gamma = -\ln P(\tau)/\tau$ is the effective decay rate. In more general situation with stochastic measurement protocol, the survival probability after $N$ measurements is given by $\prod_{i=1}^{N} P(T_i)$. Due to the law of large numbers, for $N \gg 1$ this probability is equal to $e^{-\Gamma t}$, where $t = N \langle T \rangle$ and
\begin{equation}
 \Gamma=-\frac{\int_0^\infty dT \rho(T)\ln P(T)}{\int_0^\infty dT \rho(T)T}=-\frac{\langle\ln P(T)\rangle}{\langle T\rangle}.
\end{equation}
When the measurement frequency is high enough, deviation of $P(T)$ from unity for typical values of $T$ is small (this is widely used assumption in the literature) and we can safely replace $\langle \ln P(T) \rangle \approx 1 - \langle P(T) \rangle$. Then, using Eq.~(\ref{stochastic}), one obtains $\Gamma \approx \langle 1 - P(T) \rangle/\langle T \rangle = \langle t \rangle^{-1}$.

%

\begin{figure}[t]
  \includegraphics[width=0.65\linewidth]{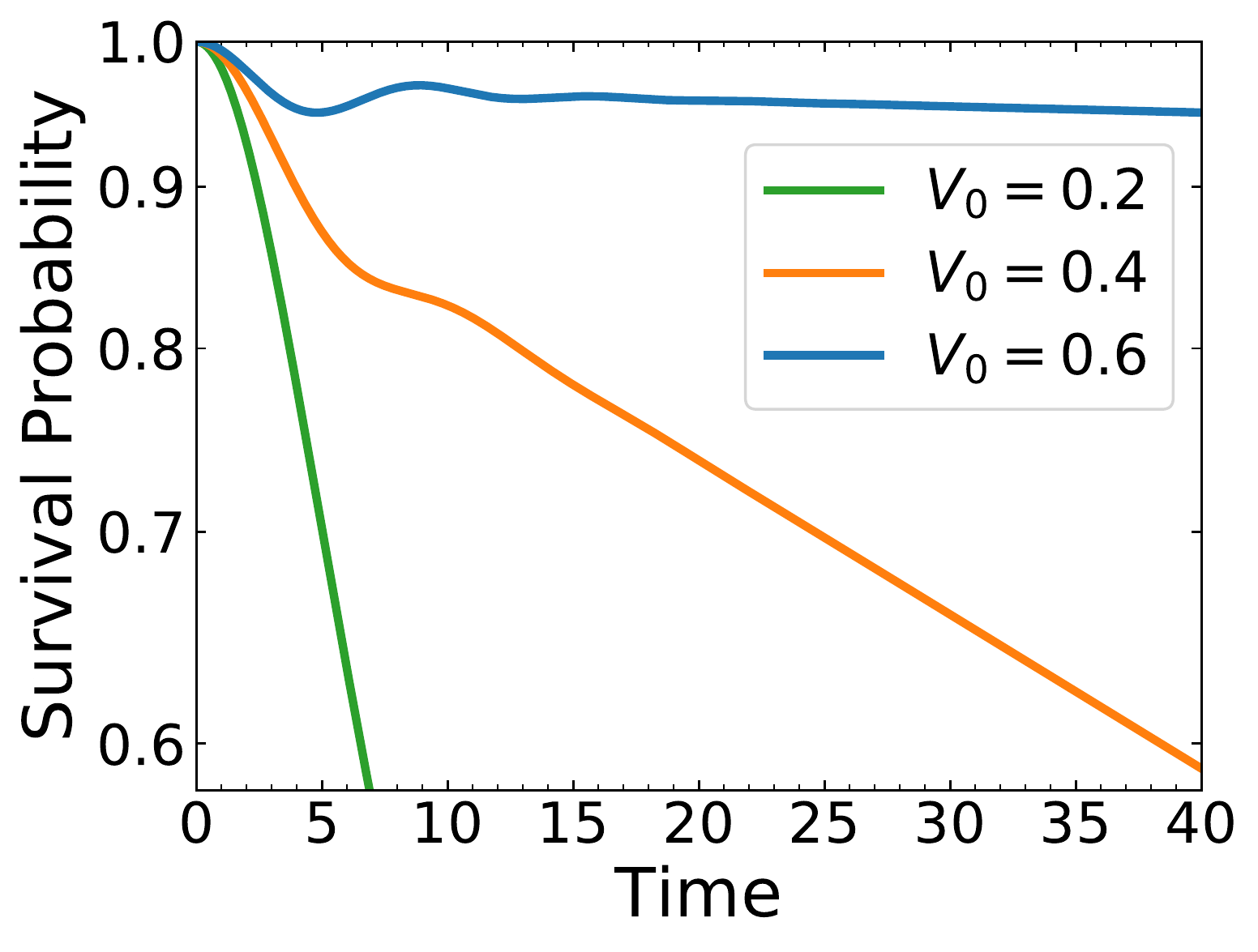}
  \captionsetup{format=plain, justification=centerlast}
  \caption{The probability $P(t)$ of finding the trapped atom in its initial state as a function of time $t$ for different potential amplitudes $V_0$. The acceleration of optical potential is equal to $a=0.1$.}
\label{fig:1}
\end{figure}

\textbf{Escape probability of trapped atoms.}
To obtain a simple expression for the tunnelling probability, one can keep only the gap between the first two Bloch bands, and neglect all the other bandgaps. This leads to a modified band structure with a single band, separated by a bandgap from free particle motion. This approximation is valid when the tunnelling rate across the higher gaps is sufficiently high. By treating the non-adiabatic coupling between the trapped and non-trapped states as a weak perturbation, one can find that to leading order the logarithm of the survival probability in the trapped state is equal to \cite{Wilkinson_1997}
\begin{equation}
\label{eq:surv_prob}
\ln P(t) = - \int_{0}^{t} d \tau (t-\tau) W(\tau),
\end{equation}
where
\begin{eqnarray}
\nonumber
   W(\tau) = &\displaystyle \dfrac{a^2}{2 V_0} \int_{-\infty}^{\infty} ds \dfrac{1}{1+(s-a\tau/V_0)^2} \dfrac{1}{1+s^2}&\\
   &\displaystyle \times \cos \left( \dfrac{V_0^2}{a} \int_{s}^{s-a\tau/V_0} dz \sqrt{1+z^2} \right),&
\end{eqnarray}
and here we use dimensionless units introduced in the main text. In the parameter space of $a$ and $V_0$, the theory is valid inside the region bounded by the two curves $a=\pi V_0^2/2$ and $a = \pi V_0^4/4$, and to the left of the line $V_0=1$ \cite{Niu_1998}. The presented expression demonstrates good agreement with experimental results and captures a short-term deviation from the exponential decay, see Fig.~\ref{fig:1}.

\textbf{Proof of the optimality of the stroboscopic protocol.}
Assume that $\tau^*$ is the optimal period of deterministic measurement protocol minimizing the mean time to decay which is given  by Eq. (\ref{stroboscopic}), i.e.
\begin{equation}
\label{determ_1}
\frac{\tau_\ast}{1-P(\tau_\ast)}\le \frac{\tau}{1-P(\tau)},
\end{equation}
for all $0<\tau<\infty$. Below we prove that for any measurement procedure $\{T_k\}_{k=1}^{+\infty}=T_1,T_2,\dots$ the resulting expected decay time $\langle t\rangle$ is larger than $\langle t_{\tau_\ast}\rangle$. From Eq.~(\ref{decay_time}) we find
\begin{eqnarray*}
\langle t\rangle=T_1 + \sum_{n=2}^{\infty} T_n\prod_{k=1}^{n-1}P(T_k)=\frac{T_1}{1-P(T_1)}(1-P(T_1)) +\\
+ \sum_{n=2}^{\infty} \frac{T_n}{1-P(T_n)}(1-P(T_n))\prod_{k=1}^{n-1}P(T_k).
\end{eqnarray*}
Next, we note that $1-P(T_1)$ is the probability to register the decay of unstable system after the first measurement $Q_1$, and accordingly, $Q_n = (1-P(T_n))\prod_{k=1}^{n-1} P(T_k)$ is the probability of registering the decay after the $n$-th measurement. Also, we remind that $T_n/(1-P(T_n))=\langle t_{T_n}\rangle$ is the expected decay time under stroboscopic protocol of period $T_n$, and thus we obtain
\begin{equation}
\label{optimality}
\langle t\rangle=\sum_{n=1}^{\infty} \langle t_{T_n}\rangle Q_n.
\end{equation}
Further we exploit the facts that $\langle t_{T_n}\rangle \ge \langle t_{\tau_\ast}\rangle$ (direct consequence of Eq. (\ref{determ_1}) and $\sum_{n=1}^{\infty} Q_n=1$ (normalization condition) to obtain from Eq.~(\ref{optimality})
\begin{equation}
\langle t\rangle\ge\langle t_{\tau_\ast}\rangle.
\end{equation}
This proves that deterministic measurement protocol is optimal  among all possible strategies.
Note that the above arguments do not invoke any system-dependent features and thus our conclusion is universally valid for any unstable system.


\textbf{Upper limit for the decay time in Luby-like protocol.}
In our derivation of Eq. (\ref{Luby_bound}) we follow the line of argumentation proposed in Ref.~\cite{Luby_1993} in the context of probabilistic algorithms under restart.

For any $j$, if the total time spent
on inter-measurement runs of length $2^j \tau_0$ up to the end of some run in the
sequence is $W$, then at most $(\log_2 W/\tau_0 + 1)$ different
inter-measurements intervals have so far been used, and the total time
spent on each one cannot exceed $2W$.
Thus the total time elapsed since the start of the experiment
 up to this point  is at
most $2W(\log_2 W/\tau_0 + 1)$.

As above, we denote as $\tau^* \gg \tau_0$ the optimal period of the stroboscopic measurement protocol minimizing the decay time given  by Eq. (\ref{stroboscopic}).
We set $i_0 = \lceil \log_2(\tau_\ast/\tau_0) \rceil$ and
$m_0 = \lceil \log_2(1/Q(\tau_\ast)) \rceil$, where $Q(T)=1-P(T)$ and $\lceil \dots \rceil$ denotes the procedure of rounding up to an integer. Consider the instant when $2^{m_0}$
runs of length $2^{i_0}\tau_0$ have been executed. The probability
that the system has failed to decay on all of these runs is at
most
\begin{equation}
(1-Q(2^{i_0}\tau_0))^{2^{m_0}}\le (1-Q(\tau_\ast))^{1/Q(\tau_\ast)}\le e^{-1},
\end{equation}
where we have used an assumption that $P(T)$ is monotonically decreasing function.
At this point, the total time spent on runs of length
$2^{i_0}\tau_0$ is
\begin{equation}
\label{eq:Lb}
W=2^{i_0+m_0}\tau_0\le 4 \langle t_{\tau_\ast} \rangle,
\end{equation}
due to Eq.~(\ref{stroboscopic}), and by the observation above the total time spent up
to this point is at most $2W(\log_2 W/\tau_0 + 1)$. More generally,
after $k2^{m_0}$ runs of length $2^{i_0}\tau_0$ have been completed,
the probability that the system has failed to decay is at
most $e^{-k}$, and the total time spent up to this point is
at most $2kW(\log_2(kW/\tau_0) + 1)$. Therefore,
\begin{equation}
\langle t_{univ}\rangle\le \sum_{k=1}^{\infty}2kW(\log_2(kW/\tau_0) + 1)e^{-k+1}.
\end{equation}
Taking into account Eq.~(\ref{eq:Lb}) and relation $\sum_{k=1}^\infty ke^{-k+1}=\frac{1}{(1-e^{-1})^2}$ we readily obtain Eq.~(\ref{Luby_bound}).

\textbf{Non-uniform scale-free protocol.}
Let us assume that the survival probability has the form $P(T)=f(T/gT_0)$, where $T_0$ is the characteristic time associated with quantum evolution, and the dimensionless constant $g$ represents the scale factor.
We implement to this process the randomly distributed measurements at a non-uniform  rate  which  is  inversely  proportional  to  the  time elapsed  since  the  start  of  the  experiment, i.e. $r(t)=\alpha/t$, where $\alpha$ is a dimensionless constant.
This protocol is characterized by random scale-free time intervals between successive measurements.
Namely, given a measurement at time $t$, the time till next measurement has the following distribution (see \cite{Kusmierz_2018})
\begin{equation}
\rho(\tau|t)=\frac{\alpha t^{\alpha}}{(\tau+t)^{\alpha+1}},
\end{equation}
which implies that $\tau(t)$ has the same distribution as $g\tau(t/g)$.
The remaining time to decay satisfies the following renewal equation
\begin{equation}
T_\alpha(t,g)=\tau(t)+T'_\alpha(t+\tau(t),g)x_{\frac{\tau(t)}{gT_0}}
\end{equation}
Using the observation above, we can rewrite this equation in the form
\begin{equation}
T_\alpha(gt,g)=g\tau(t)+T'_\alpha(gt+g\tau(t),g)x_{\frac{\tau(t)}{T_0}}
\end{equation}
Thus, we see that
\begin{equation}
T_\alpha(gt,g)\sim gT_\alpha(t,1),
\end{equation}
where the symbol $\sim$ means that the random variables have the same distribution.
Taking the limit $t \to 0$, we conclude that the decay time of the system subject to the scale-free measurements
scales linearly with the characteristic time scale of the underlying quantum evolution.
Therefore, the mean value is proportional to $g$.
Due to this feature, the optimal parameter $\alpha_\ast$, which brings expected decay time to a minimum, is insensitive to $g$.
This allow us to guarantee the optimal performance in the absence of knowledge about exact value $g$.

\textbf{Poisson Measurements.}
For any measurement procedure which is uniform in time, the expected time to decay satisfies the following renewal equation
\begin{equation}
t=T+t'x_T,
\end{equation}
where $x_T$ is the binary random variable which is equal to zero if the outcome of the current measurement is "decay" and is unity otherwise, $T$ is the random time between measurements sampled from $\rho(T)$, and $t'$ is independent copy of $t$.
Therefore
\begin{equation}
t^2=T^2+2t'x_TT+t'^2x_T,
\end{equation}
After averaging over the statistics of measurements and quantum statistics we obtain
\begin{equation}
\langle t^2\rangle=\langle T^2 \rangle+2\langle t' x_TT\rangle+\langle t'^2x_T\rangle,
\end{equation}
Since $\langle t' x_TT\rangle=\langle t'\rangle \langle x_TT\rangle$, $\langle t'^2x_T\rangle=\langle t'^2\rangle\langle x_T\rangle$, $\langle t\rangle=\langle t'\rangle$ and $\langle t^2\rangle=\langle t'^2\rangle$, one obtains
\begin{equation}
\label{tmp}
\langle t^2\rangle=\frac{\langle T^2 \rangle+2\langle t\rangle\langle x_TT\rangle}{1-\langle x_T\rangle}.
\end{equation}
If the measurement events come from Poisson statistics, then $\langle T^2 \rangle=2r^{-2}$, $\langle x_T\rangle=r\int_0^{\infty}dT P(T)e^{-rT}=r\tilde{P}(r)$, $\langle x_TT\rangle=r\int_0^{\infty}dT TP(T)e^{-rT}=-r\partial_r\tilde{P}(r)$, and $\langle t\rangle$ is given by Eq.~(\ref{Poisson}).
Substituting these expressions into Eq.~(\ref{tmp}) gives Eq.~(\ref{Poisson2}).

Next, using Eq.~(\ref{Poisson}), we can express the Laplace transform $\tilde{P}(r)$ and its derivative $\partial_r\tilde{P}(r)$ through $r$ and $\langle t_r\rangle$
\begin{eqnarray}
\label{eq:1}
&\displaystyle \tilde{P}(r)=\frac{1}{r}-\frac{1}{r^2\langle t_r\rangle},&\\
\label{eq:2}
&\displaystyle \partial_r\tilde{P}(r)=-\frac{1}{r^2}+\frac{2}{r^3\langle t_r\rangle}+\frac{\partial_r \langle t_r\rangle}{r^2\langle t_r\rangle^2}.&
\end{eqnarray}
It follows from Eq.~(\ref{eq:2}) that if we found an optimal measurement rate $r^\ast$ such that the expected decay time attains its minimum value, i.e. $\partial_r \langle t_{r_\ast}\rangle=0$, then $\partial_r\tilde{P}(r_\ast)=-\frac{1}{r_\ast^2}+\frac{2}{r^3_\ast\langle t_{r_\ast}\rangle}$ and substituting of this relation together with Eq.~(\ref{eq:1}) into Eq.~(\ref{Poisson2}) yields Eq.~(\ref{Poisson_universality}).

To derive Eq.~(\ref{Poisson_universality_2}) let us note that $\partial_r\langle t_r\rangle=\frac{-2\tilde{Q}(r)-r\partial_r\tilde{Q}(r)}{r^3\tilde{Q}^2(r)}$ where $\tilde{Q}(r)$ is the Laplace transform of the function $Q(T)=1-P(T)$ evaluated at $r$.
At $r=r_\ast$ one has $\partial_r\langle t_{r_\ast}\rangle=0$ so that $-\frac{\partial_r\tilde{Q}(r_\ast)}{\tilde{Q}(r_\ast)}=2/r_\ast$.
Introducing the time $\tilde{T}_r=\frac{\langle (1-x_T)T\rangle}{1-\langle x_T\rangle}=\frac{\int_0^{\infty}dT TQ(T)e^{-rT}}{\int_0^{\infty}dT Q(T)e^{-rT}}=-\frac{\partial_r\tilde{Q}(r)}{\tilde{Q}(r)}$ we arrive at Eq.~(\ref{Poisson_universality_2}).

\textbf{Numerical simulations.}
We use different numerical methods for the deterministic and stochastic measurement strategies. The simulations in the deterministic case were based on Eq.~(\ref{decay_time}) and we interrupted the numerical summation over $n$ when the survival probability $\prod_{k=1}^{n-1}P(T_k)$ became less than $\epsilon=10^{-10}$. We checked that a further increase in accuracy practically does not change the final results. In the stochastic case we simulated the behaviour of ensemble of trapped atoms and collected statistics from at least $N=10^6$ independent atoms.

%

\textbf{Author Information.}
S.B. initiated the study and derived the analytical results. V.P. performed numerical simulations and analysed the data. Both authors were involved in drafting the final manuscript.

\textbf{Acknowledgments.}
S.B. gratefully acknowledges support from the James S. McDonnell Foundation via its postdoctoral fellowship in studying
complex systems.

\textbf{Competing interests.}
The authors declare no competing financial interests.


{}


\begin{thebibliography}{99}


\bibitem{Hodges} A. Hodges, "Alan Turing: Life and Legacy of a Great
Thinker", p.54

\bibitem{Misra_1977} B. Misra and E.C.G. Sudarshan, J. Math. Phys. 18, 756
(1977)

\bibitem{Itano_1990} W.M. Itano, D.J. Heinzen, J.J. Bollinger, and D.J.
Wineland, Phys. Rev. A 41, 2295 (1990).

\bibitem{Kwiat_1995} P. Kwiat, H. Weinfurter, T. Herzog, A. Zeilinger, and
M.A. Kasevich, Phys. Rev. Lett. 74, 4763 (1995).

\bibitem{Streed_2006} E. W. Streed, J. Mun, M. Boyd, G. K. Campbell, P. Medley, W. Ketterle, and D. E. Pritchard,
Continuous and Pulsed Quantum Zeno Effect, Phys. Rev. Lett. 97, 260402 (2006).

\bibitem{Helmer_2009} F. Helmer, M. Mariantoni, E. Solano, and F. Marquardt,
Quantum nondemolition photon detection in circuit QED and the quantum Zeno effect, Phys. Rev. A 79, 052115 (2009).

\bibitem{Wolters_2013} J. Wolters, M. Strauß, R. S. Schoenfeld, and O. Benson, Quantum Zeno phenomenon on a single solid-state spin, Phys. Rev. A 88, 020101 (2013).
    
\bibitem{Patil_2015} Y. S. Patil, S. Chakram, and M. Vengalattore, Phys. Rev. Lett. 115, 140402 (2015).


\bibitem{Kakuyanagi_2015} K. Kakuyanagi, T. Baba, Y. Matsuzaki, H. Nakano, S. Saito, and K. Semba, Observation of quantum Zeno effect in a superconducting flux qubit, New J. Phys. 17, 063035 (2015).
    
\bibitem{Slichter_2016} D. H. Slichter, C. M¨uller, R. Vijay, S. J. Weber, A. Blais, and I. Siddiqi, Quantum Zeno effect in the strong measurement regime of circuit quantum electrodynamics, New J. Phys. 18, 053031 (2016).
    
\bibitem{Kalb_2016} N. Kalb, J. Cramer, D. J. Twitchen, M. Markham, R. Hanson, and T. H. Taminiau, Experimental creation
of quantum Zeno subspaces by repeated multi-spin projections in diamond, Nat. Commun. 7, 13111 (2016).
    

\bibitem{Fischer_2001} M.C. Fischer, B. Gutierrez-Medina, and M.G. Raizen,
Phys. Rev. Lett. 87, 040402 (2001).


\bibitem{Harrington_2017} P. M. Harrington, J. T. Monroe, and K. W. Murch, Quantum Zeno Effects from Measurement Controlled Qubit-Bath Interactions, Phys. Rev. Lett. 118, 240401 (2017).


\bibitem{Kofman_2000} A. G. Kofman and G. Kurizki, Acceleration of quantum decay processes by frequent observations, Nature 405, 546 (2000).

\bibitem{Facchi_2001} Facchi, P and Nakazato, H and Pascazio, S, From the quantum Zeno to the inverse quantum Zeno effect, Physical Review Letters 86, 2699 (2001).

\bibitem{Koshino_2005} Koshino, K., and Shimizu, A. (2005). Quantum Zeno effect by general measurements. Physics reports, 412(4), 191-275.


\bibitem{Ghirardi_1979} Ghirardi, G. C. and Omero, C. and Weber, Tullio and Rimini, Alberto. Small-time behaviour of quantum nondecay probability and Zeno's paradox in quantum mechanics. Il Nuovo Cimento A, 52(4), 421--442 (1979).









\bibitem{Barone_2004} A. Barone, G. Kurizki, and A. G. Kofman, Dynamical Control of Macroscopic Quantum Tunneling,
Phys. Rev. Lett. 92, 200403 (2004).

\bibitem{Chen_2010} P.-W. Chen, D.-B. Tsai, and P. Bennett, Quantum Zeno and anti-Zeno effect of a nanomechanical resonator measured by a point contact, Phys. Rev. B 81, 115307 (2010).















\bibitem{Segal_2007} Segal, D. and Reichman, D. R. Zeno and anti-Zeno effects in spin-bath models. Phys. Rev. A 76, 012109 (2007).

\bibitem{Thilagam_2010}  Thilagam, A. Zeno-anti-Zeno crossover dynamics in a spin-boson system. J. Phys. A: Math. Theor. 43, 155301 (2010).

\bibitem{Chaudhry_2014} Chaudhry, A. Z. and Gong, J. Zeno and anti-Zeno effects on dephasing. Phys. Rev. A 90, 012101 (2014).

\bibitem{Chaudhry_2016} A. Z. Chaudhry,  A general framework for the quantum Zeno and anti-Zeno effects, Scientific reports 6, 29497 (2016).

\bibitem{Aftab_2017} Aftab, M. J., and Chaudhry, A. Z. (2017). Analyzing the Quantum Zeno and anti-Zeno effects using optimal projective measurements. Scientific reports, 7(1), 11766.

\bibitem{Chaudhry_2017} Chaudhry, A. Z. (2017). The quantum Zeno and anti-Zeno effects with strong system-environment coupling. Scientific reports, 7(1), 1741.

\bibitem{Luby_1993} M. Luby, A. Sinclair, and D. Zuckerman, Optimal speedup of Las Vegas algorithms,  Inf. Proc. Lett. {\bf{47(4)}}, 173-180 (1993).

\bibitem{Kusmierz_2018} Kusmierz, L., and Toyoizumi, T. (2018). Robust parsimonious search with scale-free stochastic resetting. arXiv preprint arXiv:1812.11577.

\bibitem{Wilkinson_1997} Wilkinson, S. R. and Bharucha, C. F. and Fischer, M. C. and Madison, K. W. and Morrow, P. R. and Niu, Q. and Sundaram, B. and Raizen, M. G., Experimental evidence for non-exponential decay in quantum tunnelling, Nature 387, 575 (1997).



\bibitem{Facchi_2008} P. Facchi and S. Pascazio, Quantum Zeno dy-
namics: Mathematical and physical aspects,
J. Phys. A: Math. Theor. 41, 493001 (2008).

\bibitem{Zhang_2018} Zhang, J. M., Jing, J., Wang, L. G., and Zhu, S. Y. (2018). Criterion for quantum Zeno and anti-Zeno effects. Physical Review A, 98(1), 012135.






\bibitem{Krovi_2006} Krovi, H., and Brun, T. A. (2006). Quantum walks with infinite hitting times. Physical Review A, 74(4), 042334.

\bibitem{Varbanov_2008} Varbanov, M., Krovi, H., and Brun, T. A. (2008). Hitting time for the continuous quantum walk. Physical Review A, 78(2), 022324.

\bibitem{Stefanak_2008} Štefaňák, M., Jex, I., and Kiss, T. (2008). Recurrence and Pólya number of quantum walks. Physical review letters, 100(2), 020501.

\bibitem{Darazs_2010} Darázs, Z., and Kiss, T. (2010). Pólya number of the continuous-time quantum walks. Physical Review A, 81(6), 062319.

\bibitem{Krapivsky_2014} Krapivsky, P. L., Luck, J. M., and Mallick, K. (2014). Survival of classical and quantum particles in the presence of traps. Journal of Statistical Physics, 154(6), 1430-1460.

\bibitem{Dhar_2015a} Dhar, S., Dasgupta, S., and Dhar, A. (2015). Quantum time of arrival distribution in a simple lattice model. Journal of Physics A: Mathematical and Theoretical, 48(11), 115304.

\bibitem{Dhar_2015b} Dhar, S., Dasgupta, S., Dhar, A., and Sen, D. (2015). Detection of a quantum particle on a lattice under repeated projective measurements. Physical Review A, 91(6), 062115.

\bibitem{Friedman_2017} Friedman, H., Kessler, D. A., and Barkai, E. (2017). Quantum walks: The first detected passage time problem. Physical Review E, 95(3), 032141.


\bibitem{Thiel_2018a} Thiel, F., Barkai, E., and Kessler, D. A. (2018). First detected arrival of a quantum walker on an infinite line. Physical review letters, 120(4), 040502.

\bibitem{Thiel_2018b} Thiel, F., Kessler, D. A., and Barkai, E. (2018). Spectral dimension controlling the decay of the quantum first-detection probability. Physical Review A, 97(6), 062105.

\bibitem{Lahiri_2019} Lahiri, S., and Dhar, A. (2019). Return to the origin problem for a particle on a one-dimensional lattice with quasi-Zeno dynamics. Physical Review A, 99(1), 012101.

%

\bibitem{Belan_2018} Belan, S. (2018). Restart could optimize the probability of success in a Bernoulli trial. Physical review letters, 120(8), 080601.

\bibitem{Belan_2019} Belan, S. (2019). Median and Mode in First Passage under Restart. arXiv preprint arXiv:1906.05619.

\bibitem{Roldan_PRE_2017}  {\'E}. Rold{\'a}n,  A. Lisica, D. S{\'a}nchez-Taltavull, and S. W. Grill,  Stochastic resetting in backtrack recovery by RNA polymerases, Phys. Rev. E {\bf{93(6)}}, 062411 (2016).

\bibitem{Montanari_2002} A. Montanari, and R. Zecchina,  Optimizing searches via rare events, Phys. Rev. Lett. {\bf{88(17)}}, 178701 (2002).


\bibitem{EM_2011}  M.R. Evans and S.N. Majumdar, Diffusion with stochastic resetting, Phys. Rev. Lett. {\bf{106}},
160601 (2011).

\bibitem{Evans_2011} M. R. Evans and S. N. Majumdar,  Diffusion with optimal resetting,  J. Phys. A {\bf{44(43)}}, 435001 (2011).


\bibitem{Evans_2014} M. R. Evans and S. N.  Majumdar, Diffusion with resetting in arbitrary spatial dimension,  J. Phys. A {\bf{47(28)}}, 285001 (2014).

\bibitem{Kusmierz_2014} {\L}. Ku{\'s}mierz, S. N. Majumdar, S. Sabhapandit, and
G. Schehr,   First order transition for the optimal
search time of L{\' e}vy flights with resetting, Phys. Rev.
Lett. {\bf{113(22)}}, 220602 (2014).

\bibitem{Kusmierz_2015} {\L}. Ku{\'s}mierz and E. Gudowska-Nowak,   Optimal first-arrival times in L{\' e}vy flights with resetting, Phys. Rev. E {\bf{92(5)}}, 052127 (2015).



\bibitem{Pal_2016} A. Pal, A. Kundu, and M.R. Evans, Diffusion
under time-dependent resetting,  J. Phys. A {\bf{49(22)}}, 225001 (2016).

\bibitem{Eule_2016} S. Eule, and J. J. Metzger,  Non-equilibrium steady states of stochastic processes with intermittent resetting, New Journal of Physics {\bf{18(3)}}, 033006 (2016).

\bibitem{Nagar_2016} A. Nagar and S. Gupta,  Diffusion with stochastic resetting at power-law times, Phys. Rev. E {\bf{93(6)}}, 060102 (2016).

\bibitem{Rotbart_2015} T. Rotbart, S. Reuveni, and M. Urbakh,  Michaelis-Menten reaction scheme as a unified approach towards the optimal restart problem, Phys. Rev. E {\bf{ 92(6)}}, 060101 (2015).

\bibitem{Reuveni_PRL_2016} S. Reuveni,   Optimal stochastic restart renders fluctuations in first passage times universal, Phys. Rev. Let. {\bf{116(17)}}, 170601 (2016).

\bibitem{Reuveni_PRL_2017} A. Pal, and S. Reuveni, First Passage under Restart, Phys. Rev. Lett. {\bf{118(3)}}, 030603 (2017).


%
%
%
%
%
%
%
%
%
%
%
%
%
%
%
%
%
%
%
%
%
%
%
%
%
%
%
%
%
%
%
%
%
%
%
%
%
%
%
%
%
%
%



%
%
%
%




%

%
%
%

%
%
%
%
%
%
%



%
%


%
%
%
%
%
%
%


%
%
%
%


%
%

\bibitem{Belan_2018} S. Belan, Restart could optimize the probability of success in a Bernoulli trial, Physical review letters 120(8), 080601 (2018)

\bibitem{Chechkin_2018} Chechkin, A., and Sokolov, I. M. (2018). Random search with resetting: a unified renewal approach. Physical review letters, 121(5), 050601.

\bibitem{Pal_2019} Pal, A., Eliazar, I., and Reuveni, S. (2019). First passage under restart with branching. Physical review letters, 122(2), 020602.


\bibitem{Power_1996} W. I. Power and P. L. Knight, Phys. Rev. A 53, 1052 (1996).

\bibitem{Belavkin_2000} V. P. Belavkin and P. Staszewski, J. Math. Phys. 41, 7220 (2000).

\bibitem{Shushin_2011} Shushin, A. I. (2011). The effect of measurements, randomly distributed in time, on quantum systems: stochastic quantum Zeno effect. Journal of Physics A: Mathematical and Theoretical, 44(5), 055303.

\bibitem{Gherardini_2016} Gherardini, S., Gupta, S., Cataliotti, F. S., Smerzi, A., Caruso, F., and Ruffo, S. (2016). Stochastic quantum Zeno by large deviation theory. New Journal of Physics, 18(1), 013048.

\bibitem{Muller_2017} Müller, M. M., Gherardini, S., and Caruso, F. (2017). Quantum Zeno dynamics through stochastic protocols. Annalen der Physik, 529(9), 1600206.

\bibitem{Niu_1998} Niu, Qian and Raizen, M. G. (1998). How Landau-Zener tunneling takes time. Physical Review Letters 80, 3491.



%
%





\end{thebibliography}
\end{document}